\documentclass[12pt,preprint]{aastex}

\begin{document}

\title {DIRECT DETECTION OF EXTRA-SOLAR COMETS IS POSSIBLE}

\author{M. Jura} 
\affil{Department of Physics and Astronomy, University of California,
    Los Angeles CA 90095-1562; jura@clotho.astro.ucla.edu}

\begin{abstract}
     
 The dust tails of comets similar to  Hale-Bopp  can
 scatter as much optical sunlight as does the Earth.  Space-based observatories such as the Terrestrial Planet Finder or ${\it Darwin}$  that will detect
extra-solar terrestrial planets also  will be able to detect extra-solar comets.

\end{abstract}
\section{INTRODUCTION}
Bright comets have been studied for centuries and are endlessly fascinating.  However,   we do not  know whether
comets in the Solar System are typical or freakish.   Here, we show that  planned space missions  such
as TPF-C (Terrestrial Planet Finder -- Coronagraph) or ${\it Darwin}$ 
that are being designed to detect extra-solar terrestrial planets also will be able to detect  extra-solar comets and thus
test models for their formation and evolution.  

Comets may have been  indirectly detected around a variety of stars.  
Time-varying narrow absorption lines have been attributed to cometary tails  around  12 Myr-old  ${\beta}$ Pic
(Artymowicz 1997, Vidal-Madjar et al. 1998, Zuckerman 2001), and other young stars (Grady et al. 1996, Roberge et al. 2002).  However, although Lecavelier des Etangs et al. (1996) and Li \& Greenberg (1998) have proposed that much of
the dust around ${\beta}$ Pic arises from comets, the observed  transient gas-phase absorbers contain a substantial amount of refractory material, and the degree of  resemblance of these parent-bodies to ice-rich comets in the Solar System is  uncertain (Karmann et al. 2001, 2003, Thebault et al. 2003).

 Ensembles of comets
around high-luminosity red giant stars where ice is rapidly sublimated (Stern et al. al. 1990, Ford \& Neufeld 2001, Ford et al. 2004) might exist, and such a system could
 explain the  detection of gaseous H$_{2}$O in the outflow from IRC+10216, a carbon-rich mass-losing red giant (Melnick et al. 2001). Alternatively, however, Willacy (2004) has proposed that the 
observed H$_{2}$O results from catalysed reactions on the surfaces of dust grains carried in the stellar wind and does not signify a population of orbiting comets.  Jura (2004) has argued that the lack
of excess 25 ${\mu}$m radiation around first ascent red giants means that
these stars typically have less than 0.1 M$_{\oplus}$ of comet-like Kuiper Belt Objects in orbits at ${\sim}$45 AU.

Comets might be indirectly detected around degenerate stars.  
Cometary impacts can explain the presence
of metals in the atmospheres of white dwarf stars which otherwise would be pure hydrogen or pure helium (Alcock et al. 1986), but this hypothesis is quite uncertain (see Zuckerman et al. 2003).   Collisions of comets with neutron stars could produce  ${\gamma}$-ray bursts, and upper limits to the rate of  
such  bursts   constrain the 
 frequency of comets in orbit around neutron stars  (Shull \& Stern 1995).  

With high-resolution ground-based optical observations, it is possible to detect individual comet gas-tails around solar-type main-sequence stars through transient absorption of the OH bands near 3100  {\AA}.  For extra-solar environments similar to the Solar System, averaged over all possible inclinations, the chance
in any randomly timed observation of detecting a large comet like Hale-Bopp by this method is only 3 ${\times}$ 10$^{-8}$, but around young stars
with infrared excesses, this probability may approach 0.01 (Jura 2005). 

All of these methods are somewhat indirect and are yet to provide compelling
evidence for extra-solar comets.
Here, we describe how comets might be directly detected because of scattering by dust in their extended tails. 
In particular,  although the final system architecture is not developed, TPF-C is currently envisioned
as a 4 by 6 m optical telescope with a coronagraph designed to find the scattered light from Earth-like
planets in the habitable zone around main-sequence stars within ${\sim}$ 15 pc of the Sun.  
We argue that any system which will be able to obtain
direct images of an analog to the Earth also will be able to detect bright comets. 

\section{DUST SCATTERING IN COMET TAILS}

\subsection{Overview}

We consider only the portion of a comet's orbit where it is close enough to its host star that 
there is rapid water-ice sublimation and consequently a large release of dust embedded within this ice.  By integrating over the entire portion of the orbit where the comet is vigorously outgassing, we  then
compute the total mass of dust ejected by the comet.  We show that most of the dust approximately follows the orbital trajectory of the
comet's nucleus, and we  estimate the amount of light scattered by the comet's dust tail.

We denote   the water ejected by a comet between two times, $t_{1}$ and $t_{2}$, as
 ${\Delta}M_{H_{2}O}(t_{1},t_{2})$ and we set $t$ = 0 at the time
of the comet's periastron.  If ${\chi}_{eff}$ (cm$^{2}$ g$^{-1}$) denotes
the effective scattering opacity of the dust associated with the ejected water, then in the optically thin tail, the fraction, $f_{comet}$, of the host star's luminosity scattered
by a comet at distance $R$ from the host star is:
\begin{equation}
f_{comet}\;{\approx}\;\frac{{\chi}_{eff}{\Delta}M_{H_{2}O}(t_{1},t_{2})}{4{\pi}R^{2}}
\end{equation}

We apply equation (1) to an extra-solar analog to a large comet such as
 Hale-Bopp  with a radius over 20 km and a total mass of  ${\sim}$ 3 ${\times}$ 10$^{19}$ g.  Comets might eject 10$^{-4}$ of their total mass during a passage near the Sun (see, for example, Whipple \& Huebner 1976).  Thus,  an analog to Hale-Bopp could release  3 ${\times}$ 10$^{15}$ g.  As noted in $\S2.2$, a typical
scattering opacity for dust in Solar System comets is  ${\chi}_{eff}$ = 160 cm$^{2}$ g$^{-1}$.  
In this  case, with $R$ = 1 AU, then
$f_{comet}$  = 1.7 ${\times}$ 10$^{-10}$.  Adopting 0.37 as the Earth's albedo (Tholen et al. 2000), then even at the most favorable phase function,
the fraction of  the Sun's luminosity scattered by the Earth  is 1.4 ${\times}$ 10$^{-10}$.  Thus, any optical facility designed to discover an analog to the Earth by its scattered light also may discover scattered light from bright comets.   

\subsection{Detailed Standard Model} 
We now perform a more exact 
evaluation of  equation (1); we need to estimate both ${\chi}_{eff}$ and ${\Delta}M_{H_{2}O}(t_{1},t_{2})$.  
We derive ${\chi}_{eff}$ using data from  A'Hearn et al. (1995) who observed 85 comets between 1976 and 1992 and reported $Af{\rho}$ relative to the production rate by number of OH (${\dot N_{OH}}$),   where $A$ is the Bond albedo of the dust particles ejected from the comet, and where $f$ is the fraction of the area of the projected telescope aperture of radius ${\rho}$ in which the dust scatters.  Introducing a constant of proportionality, $K_{0}$, we write:
\begin{equation}
A\,f\,{\rho}\;=\;K_{0}\,{\dot N_{OH}}
\end{equation} 
 Although $K_{0}$ varies by more than a factor of 10 among the different comets,  its average value is 1.5 ${\times}$ 10$^{-26}$ cm s.   

To estimate ${\chi}_{eff}$ from $Af{\rho}$, we equate  two expressions for the total effective scattering area of
the dust in the comet's tail: 
\begin{equation}
{\chi}_{eff}\;{\Delta}M_{H_{2}O}(t_{1},t_{2})\;=\;A\,f\,{\pi}{\rho}^{2}
\end{equation}
 The time for water to reach
the outer radius of the projected aperture is ($t_{2}$ - $t_{1}$) or  ${\rho}/v$ where $v$ is the outflow speed of the gas from the nucleus of the comet.  Therefore for observations of comets in the Solar System where $t_{2}$ is only slightly larger than $t_{1}$, we find:
\begin{equation}
{\Delta}M_{H_{2}O}(t_{1},t_{2}),\;=\;{\dot N_{H_{2}O}\;m_{H_{2}O}}\;\frac{{\rho}}{v}
\end{equation}
where $m_{H_{2}O}$ denotes the molecular weight of water.  
Consequently,  we can
write that:
\begin{equation}
{\chi}_{eff}\,{\dot N_{H_{2}O}}\;=\;\frac{A\,f\,{\rho}\,{\pi}\,v}{m_{H_{2}O}}
\end{equation}

We must also scale the OH and H$_{2}$O production rates.
The main OH source is
the photodissociation of water (Whipple \& Huebner 1976):
\begin{equation}
h{\nu}\;+\;H_{2}O\;{\rightarrow}\;OH\,+\,H
\end{equation}
Given the other very uncertainties,  we   
approximate (see Wu \& Chen 1993) that water is mostly destroyed by reaction (6).  Therefore
\begin{equation}
{\dot N}_{OH}\;{\approx}\;{\dot N_{H_{2}O}}
\end{equation}
From equations (2) -(7), we find:
\begin{equation}
{\chi}_{eff}\;{\approx}\;\frac{K_{0}\,{\pi}\,v}{m_{H_{2}O}}
\end{equation}
  Adopting a typical  outflow speed of 1 km s$^{-1}$ (Whipple \& Huebner 1976), then  ${\chi}_{eff}$ = 160 cm$^{2}$ g$^{-1}$.

For the purpose of studying extra-solar comets, we  estimate the total mass of ejected water, ${\Delta}M_{H_{2}O}(t_{1},t_{2})$, in  the case where $t_{2}$ $>>$ $t_{1}$.   We only include those portions of
the comet's orbit when it is warm enough that there is copious outflow of gas
and dust and therefore where sublimation of water is more important than
radiative cooling in determining the comet's surface temperature.
In this regime, we write:
\begin{equation}
{\dot M_{H_{2}O}}\;{\approx}\;\frac{L_{*}\,a\,m_{H_{2}O}}{4\,{\pi}\,R^{2}\,{\Delta}E}
\end{equation}
 where $a$ is the projected active area of
the comet which is at distance $R$ from the host star of luminosity $L_{*}$ (Jura 2005).
The energy required for the sublimation of each ice molecule, ${\Delta}E$ is
taken equal to 2 ${\times}$ 10$^{-12}$ erg (see also Sekanina 2002).  Equation (9) reproduces  within
a factor of 2 much more detailed models as long as the comet is closer than 2 AU to the Sun.    
 
 Using equation (9), we write:
\begin{equation}
{\Delta}M_{H_{2}O}(t_{1},t_{2})\;=\;{\int}_{t_{1}}^{t_{2}}{\dot M_{H_{2}O}}\,dt\;=\;\frac{L_{*}\,a\,m_{H_{2}O}}{4\,{\pi}\,{\Delta}E}{\int}_{t_{1}}^{t_{2}}\frac{dt}{R^{2}}
\end{equation}
We  evaluate the integral in equation (10) from the usual description of cometary orbits. 
If $J$ denotes the specific angular momentum of the comet, then
\begin{equation}
J\;=\;\left(R_{p}\,G\,M_{*}\,[1\,+\,{\epsilon}]\right)^{1/2}
\end{equation}
where $R_{p}$ is the periastron distance, $M_{*}$ is the mass of the host star
and ${\epsilon}$ is the eccentricity of the orbit.  If ${\phi}$ is the angle measured from the host star between the major axis of the orbital ellipse and the position vector of the comet, then the relationship between ${\phi}$ and $t$ can be derived from:
\begin{equation}
J\;=\;R^{2}\,\frac{d{\phi}}{dt} 
\end{equation}
Therefore, we write:
\begin{equation}
\frac{dt}{R^{2}}\;=\;\frac{d{\phi}}{J}
\end{equation}
For any time, $t$, we can find a corresponding angle, ${\phi}$, and for any
time interval, ${\Delta}t$, there is a corresponding angular spread of
the orbital motion, ${\Delta}{\phi}$. 
Substituting into equation (10), and letting 
 ${\Delta}{\phi}_{dust}$  denote the total angle swept out by the comet from that location where $R$ =
2 AU,
we   write from equations (10) - (13): 
\begin{equation}
{\Delta}M_{H_{2}O}(t_{1},t_{2})\;{\approx}\;\frac{L_{*}\,a\,m_{H_{2}O}\,{\Delta}{\phi}_{dust}}{4{\pi}{\Delta}E\left(R_{p}\,G\,M_{*}\,[1\,+\,{\epsilon}]\right)^{1/2}}
\end{equation}

To compute $f_{comet}$, we assume, as assessed in $\S2.3$,
that all of the dust ejected by the comet lies within the aperture of
the telescope and at the same location as the comet's nucleus.    
We therefore write from equations (1) and (14) that:
\begin{equation}
f_{comet}\;{\approx}\;\frac{K_{0}\,v\,L_{*}\,a\,{\Delta}{\phi}_{dust}}{16{\pi}R^{2}{\Delta}E\left(R_{p}\,G\,M_{*}\,[1\,+\,{\epsilon}]\right)^{1/2}}
\end{equation}
For an analog to Hale-Bopp, with ${\Delta}{\phi}_{dust}$ = ${\pi}$, $R_{p}$ = 0.91 AU, $R$ = 1 AU, $a$ = 1500 km$^{2}$ and $K_{0}$ and $v$ given above, we find that $f_{comet}$ =  1.9 ${\times}$ 10$^{-10}$, a result somewhat larger than the total amount of light scattered by  the Earth. 
 In Figure 1, we show a schematic representation of the results from
equation (15), by plotting the brightness of an analog to Hale-Bopp relative to the brightness of an analog to the Earth with the assumptions that the orbits are co-planar and  that we are looking
at the system face-on. This figure illustrates that the comet can be brighter than the planet.

Hale-Bopp was an unusually bright comet.  Because it was dynamically young, it still possessed a large amount of ice so that its ``active area" from which rapid sublimation occurred was
comparable to its geometric area. As comets age during subsequent passages near
the Sun, they  lose their volatiles and their ``active areas" can become relatively small compared to their total surface area.  

In Figure 1,  the model comet is brighter after periastron.
In contrast, observations of
Solar System comets often show comparable amounts of scattered light before
and after perihelion.  For extra-solar comets,  the amount of scattered light depends upon the total amount of ejected dust so that  $t_{2}$ $>>$ $t_{1}$. Observations of
Solar System comets are made with relatively high spatial resolution and therefore they mostly detect
dust which is  ejected just recently from the nucleus so that $t_{2}$ is only slightly greater than $t_{1}$. In this case, according to equation (9), the rate
of dust production is insensitive to whether the comet is pre- or post-periastron. 

Hale-Bopp was a particularly large comet, but its perihelion of 0.91 AU was not remarkable.   A smaller comet with a smaller perihelion also can produce a large amount of scattered light. To illustrate this effect,  we
show in Figure 2 a plot of the relative brightness of an analog to comet West with $a$ = 130 km$^{2}$ and $R_{p}$ = 0.20 AU. We selected comet West for this purpose from the 85 comets studied by A'Hearn et al. (1995) because it both had the second smallest perihelion  and  one of the largest effective areas. [A'Hearn et al. did not report data for Hale-Bopp because it entered the inner Solar System after 1992, the end of their 17 year observing span.] 
Figure 2 shows that the amount of scattered light from an analog to comet West could be  relatively large.  However, unless the angular resolution of a planet-hunting telescope is particularly good, an analog
to comet West may not be detected.

\subsection{Grain Dynamics}
Above, we have assumed that the grains follow the orbit of the comet's nucleus.
Because comets develop long tails as the ejected grains
separate from the nucleus, we now assess this approximation.

 The most important dynamical difference between a
comet and its ejected dust is the outward radiation pressure on the dust.
Consider ${\beta}$,  the ratio of radiation pressure to gravitational force on a grain. When a spherical grain's effective cross section equals its geometric cross section,
then:
\begin{equation}
{\beta}\;=\;\frac{3\,L_{*}}{16\,{\pi}\,G\,M_{*}\,{\rho}_{s}\,r_{gr}\,c}
\end{equation}
where $r_{gr}$ (${\mu}$m) denotes the grain radius, and ${\rho}_{s}$ denotes the density of the grain material.  For grains with ${\rho}_{s}$ = 3 g cm$^{-3}$ in orbit around a solar-type star, ${\beta}$ = 0.19 $r_{gr}^{-1}$

Radiation pressure acts like ``anti-gravity" and the effective inward
radial force on a grain is scaled from the true gravitational force by a factor of $(1\,-\,{\beta})$.  In Figure 3, we plot the orbits of a comet with nearly a parabolic orbit
and that of a dust particle that is ejected at periastron with ${\beta}$ = 0.2.
As long as the comet is within a distance of 3$R_{p}$ of the star, the grains and comet lie closer than 0.2$R$. The cometary tail is not likely to be resolved by a 5 meter telescope observing a star at ${\sim}$10 pc at 5000 {\AA}, and   our approximation that the grains in the tail and the comet are co-located
is appropriate for particles with ${\beta}$ $<$ 0.2 or $r_{gr}$ $>$ 1 ${\mu}$m.

We now estimate the fraction of cometary grains with $r_{gr}$ ${\geq}$ 1 ${\mu}$m.
 Harker et al. (2002)
have found that the grain size distribution for Hale-Bopp can be represented
by the function:
\begin{equation}
n(r_{gr})\,dr_{gr}\;{\propto}\;\left(1\,-\,\frac{r_{0}}{r_{gr}}\right)^{M}\,\left(\frac{r_{0}}{r_{gr}}\right)^{N}
\end{equation}
with $r_{0}$ = 0.1 ${\mu}$m, and both $M$  and $N$ near 3.5. Equation (17) requires some maximum grain size for otherwise it predicts an infinite mass in the particles.  However,  the total cross section, ${\sigma}_{tot}(r_{min})$, is independent of this maximum size, and we write  for an ensemble of  grains whose minimum size is $r_{min}$:
\begin{equation}
{\sigma}_{tot}(r_{min})\;=\;{\int}_{r_{min}}^{\infty}{\pi}r_{gr}^{2}n(r_{gr})\,dr_{gr}
\end{equation}
We plot ${\sigma}_{tot}(r_{min})/{\sigma}_{tot}(0.1\,{\mu}m)$ in Figure 4 to show how the total cross section scales downward as the minimum
grain size is increased.  
In particular, for distributions  with $r_{min}$ = 1 ${\mu}$m, the total cross section is
diminished by about a factor of 2 from the full ensemble with $r_{min}$ = 0.1 ${\mu}$m. Therefore, our estimate of
$f_{comet}$ given in equation (15) may be too large by this factor of 2.

\section{EXTENSIONS OF THE STANDARD MODEL}
 Above, we have shown how a comet like Hale-Bopp could be detected in
a ``standard model".  Here, we relax some of the assumptions that we made
to determine the detectability of a comet, and we consider some observational consequences of the models.   

\subsection{Different Orbital Elements}
In $\S2.2$, we assumed that the orbit was seen face-on.  Here, we note that
a comet like Hale-Bopp would probably be detectable regardless of the
orbital inclination.  Let the comet's motion define the $X-Y$ plane and use
standard polar coordinates  to describe the position of the comet with the host star at the origin and with the definition that ${\phi}$ = 0 at periastron.  For a comet with orbital eccentricity near 1, the position vector of the comet in the $X-Y$ plane, ${\vec R}$,  is:
\begin{equation}
{\vec R}\;{\approx}\;\frac{2\,R_{p}}{1\,+\,\cos\,{\phi}}\left({\hat x}\,\cos{\phi}\,+\,{\hat y}\,\sin{\phi}\right)
\end{equation}
We  define the unit vector to an external observer, ${\hat n}$, as:
\begin{equation}
{\hat n}\;=\;{\hat x}\cos{\theta}_{0}\cos{\phi}_{0}\;+\;{\hat y}\cos{\theta}_{0}\sin{\phi}_{0}\:+\;{\hat z}\sin{\theta}_{0}
\end{equation}
As seen by this external observer, the projected distance, $d$, between the comet and the host star is found by
\begin{equation}
d\;=\;|{\vec R}{\times}{\vec n}|\;{\approx}\;\frac{2\,R_{p}}{1\,+\,\cos\,{\phi}}\left(\sin^{2}{\theta}_{0}\;+\;\cos^{2}{\theta}_{0}\sin^{2}({\phi}\,-\,{\phi}_{0})\right)^{1/2}
\end{equation}
As shown in $\S2.2$, for an analog to Hale-Bopp, the range in values of ${\phi}$ where the comet is measurably bright is approximately  ${\pi}$.   Therefore, regardless of ${\phi}_{0}$, there is always a value of ${\phi}$ both  where the comet is bright and  where $\sin^{2}({\phi}\,-\,{\phi}_{0})$ ${\approx}$ 1.  At this particular orbital location, we see by inspection of equation (21), that  $d$ ${\geq}$ $R_{p}$.  For a star at 10 pc and $R_{p}$ = 1AU, this value of $d$ corresponds to an angular separation of 0{\farcs}1 which equals 5${\lambda}/D$ for a 5 m telescope observing at 5000 {\AA}.  The TPF coronagraph is being designed to
detect sources as faint as 10$^{-10}$ of the host star at offsets larger than  4${\lambda}/D$ (see Kasdin et al. 2004).  Therefore, any TPF-C system with the exquisite optics  to detect a terrestrial
analog at 1 AU from its host star also  will be able to detect an analog to Hale-Bopp in any orbital orientation.  

\subsection{Different Kinds of Host Stars}
 In $\S2.2$, we  discussed the detectability of comets around stars similar to the Sun.  We  now consider main-sequence stars with different masses and
luminosities. If a planet has a radius, $R_{pl}$, then the fraction of
the star's light that is reflected toward us, $f_{planet}$ is:
\begin{equation}
f_{planet}\;=\;\frac{{\pi}R_{pl}^{2}}{4{\pi}R^{2}}\,f_{sc}
\end{equation}
where $f_{sc}$  depends both upon
the albedo of the planet and its orbital orientation and phase.  Because we usually see less than the fully illuminated hemisphere of the planet, we expect that $f_{sc}$ for an analog to the Earth  typically is near 0.2.  Comparing $f_{planet}$ in equation (22) with $f_{comet}$ in equation (15),   we  find:
\begin{equation}
\frac{f_{comet}}{f_{planet}}\;=\;\frac{K_{0}\,v\,L_{*}\,a\,{\Delta}{\phi}_{dust}}{4{\pi}R_{pl}^{2}{\Delta}E\left(R_{p}\,G\,M_{*}\,[1\,+\,{\epsilon}]\right)^{1/2}\,f_{sc}}\;{\approx}\;2 \frac{L'_{*}}{(R'_{p}M'_{*})^{1/2}}
\end{equation}
where $L'_{*}$ and $M'_{*}$ are measured in solar units at $R'_{p}$ is measured in AU and we used the numerical values for the different terms in equation (23) employed above.  
We  see from expression (23) that the possibility of detecting
extra-terrestrial comets is  sensitive to the luminosity of the host star.
In particular, for stars with luminosities less than 0.5 L$_{\odot}$, corresponding to spectral types later than about G8 (see Drilling \& Landolt 2000), the cometary analogs to Hale-Bopp may not be any brighter than the analogs to the Earth.  According to Figure 1 in Kasdin et al. (2004), about 70\% of the 202 viable candidates for investigation with TPF are at least
as early as G8V with  (B-V) ${\leq}$ 0.74.  Thus, for most potential stars to be targeted with TPF-C, large comets may be as bright as an analog
to the Earth.     

\subsection{Infrared Signatures}
In $\S2.2$, we have shown that an analog to comet  Hale-Bopp can scatter as much
optical light as can an analog to the Earth.  For an object with an albedo,
$b$, the fraction of re-radiated infrared energy compared to the amount of
scattered light is $(1\,-\,b)/b$.  Comets usually have dust albedos lower than the Earth's 
typical albedo of  0.37
(see, for example, Lisse et al. 1998, 2002).  
  Consequently, following the arguments in $\S2.2$, those space missions designed
to detect terrestrial planets  in the thermal infrared, such as TPF-I, also will be able to
detect analogs to comet Hale-Bopp.  

\subsection {Optical  Emission Line Signatures} 

In the Solar System, the gaseous outflows from comets lead to strong
fluorescent emission lines.    Here,  we estimate the strength of the optical line emission from extra-solar comets by scaling from the optical emission spectrum 
of Solar System comets.    

Because they are photodissociated, the molecules in the outflow from a comet have a lifetime short compared
to the  time the comet spends near the host star.  We write ${\Delta}{\phi}_{gas}$ to denote
the angular motion of the comet as seen from the host star during the radiative lifetime of a particular gaseous molecule, ${\Delta}t_{gas}$.  
From equation (13),  then
\begin{equation}
{\Delta}{\phi}_{gas}\;=\;\frac{J}{R^{2}}{\Delta}t_{gas}
\end{equation}
We now adopt the simple approximation that in an extra-solar system comet that
at a particular location in its orbit, the line to continuum ratio, $X_{LC}$, is
given by the expression
\begin{equation}
X_{LC}(extra-solar)\;=\;\frac{{\Delta}{\phi}_{gas}}{{\Delta}{\phi}_{dust}}\,X_{LC}(Solar\;System)
\end{equation}
where ${\Delta}{\phi}_{dust}$ is  described in  $\S2.2$.  Implicit in
this formula is the prediction that because the amount of dust contributing to the scattered continuum is
relatively large, the observed line to continuum ratio
in extra-solar comets is likely to be appreciably lower than in Solar System comets.

Fink \& Hicks (1996) have presented a convenient set of low-resolution optical observations
of Solar System comets.  They report both a continuum flux at 6250 {\AA} with a bandwidth of 36 {\AA} and continuum-subtracted emission line strengths of C$_{2}$, NH$_{2}$, O I and CN in bands centered at 5520 {\AA}, 6335 {\AA}, 6300   {\AA} and 9180 {\AA}, respectively.  Here, we focus on the strongest cometary
emission lines which are the Swan bands of C$_{2}$ centered at 5520 {\AA}. 
  In the Solar System, the
characteristic lifetime of the C$_{2}$ molecule at 1 AU from the Sun is about 6 ${\times}$ 10$^{4}$ s (Fink
\& Hicks 1996).  Consequently, for a comet in an orbit with orbital eccentricity near 1.0 with a periastron of 1AU in orbit around a 1 M$_{\odot}$ main-sequence star, we expect from equations (11) and (24) that   ${\Delta}{\phi}_{gas}$ ${\approx}$ 0.02. For ${{\Delta}{\phi}_{dust}}$ ${\approx}$ 2,   we find from equation (25) that a representative line to continuum
ratio for an extra-solar comet is diminished by a factor of 100 from the
value such a comet would have during  ``instantaneous" observations performed within the Solar System.  

In the data of Fink \& Hicks (1996), there is a very large range in the observed line to continuum ratio in different comets; some objects do not exhibit any
detectable emission.   The comets with
the three strongest line fluxes measured by Fink \& Hicks (1996) were Halley, Encke and Swift-Tuttle where the ratio of the emission in the
C$_{2}$ bands to the continuum at 6250 {\AA} with the 36 {\AA} bandwidth ranged from 140 to 3000, respectively.  We
therefore expect  that bright extra-solar comets might display a line to continuum ratio in the Swan bands  of between 1.4 and
30. However, only 30\% of the 39 comets observed by Fink \& Hicks (1996) exhibit
such relatively strong C$_{2}$ emission; the others would have line to continuum ratios less than unity. 

To detect the emission line spectrum of a comet will be demanding. For a comet that has a continuum brightness of  2 ${\times}$ 10$^{-10}$  of a host star with m$_{V}$ = 6 mag, we find that in a bandwidth of 36 {\AA} observed with  a 5 m telescope that the photon arrival
rate is ${\sim}$ 0.006 s$^{-1}$.  Therefore, in a 1 hour exposure, for a comet where the C$_{2}$ line to continuum flux is 5, there would be 100 line photons incident upon the telescope.  With a very efficient spectrograph and detector, it may be
possible to measure this flux.   
 
\section{DISCUSSION}

We have shown that a comet like Hale-Bopp might be as bright as
a terrestrial planet.  We now address how frequently such bright comets might
appear and how we might identify the signal as being produced by a comet.  
  
In the Solar System, the arrival rate of comets the size of
Hale-Bopp or larger is about 7 ${\times}$ 10$^{-10}$ s$^{-1}$ (see Meech et al. 2004).  The duration of its bright phase is about  10$^{7}$ s, and  therefore, for the Solar System, the duty cycle when a comet like Hale-Bopp is evident is ${\sim}$1\%. 
The fraction of   main-sequence stars with comet clouds might be as low as 1\% or even less (Shull \& Stern 1995, Tremaine 1993).    Alternatively, however,
Heisler (1990) has argued that most new comets appear during relatively intense showers.  There may be stars which possess a comet infall rate 10 times that
in the Solar System, and these systems often might display a comet as bright
as any analog to the Earth.

Our discussion is based on extrapolating from
known comets in our own Solar System, but
 extra-solar comets may be quite different from ours. For example, although
there is a wide range, Solar System comets
have an effective dust opacity of  160 cm$^{2}$ g$^{-1}$.    
In contrast, the typical scattering opacity of interstellar dust particles is ${\sim}$ 10000 cm$^{2}$ g$^{-1}$  (see, for example, Spitzer 1978, and convert from the units described in that text).  This large difference in scattering efficiency results  from
interstellar grains being smaller and having higher albedos than cometary grains (see, for example, Lisse et al. 1998, 2004).     Because the growth of comets and comet-like bodies in young stellar objects is not well understood, it is imaginable that 
 comets  100 times  brighter than Hale-Bopp might
be discovered.  

There are   ways to distinguish between a comet and a planet.
First, comets have  nearly parabolic orbits while a planet's orbit
 has a much smaller eccentricity. 
Additionally, as shown in Figures 1 and 2, we expect that the amount of light reflected by a comet
is much brighter after periastron and thus the light curve of a comet
is quite different from that of a planet.  Finally, some comets produce
distinctive broad spectral emission features produced by molecular
fluorescence such as that in the C$_{2}$ Swan bands. Future facilities equipped with energy-sensitive photon detectors (see Sparks \& Ford 2002, Day et al. 2003) could efficiently achieve low-resolution spectroscopy and 
in some instances  distinguish 
between planets and comets.

\section{CONCLUSIONS}
We find that comets as large as Hale-Bopp can scatter as much optical light
as would an  analog to the Earth. Future space missions such as TPF-C or ${ \it Darwin}$ may detect large comets around solar-type stars.    If extra-solar comets eject dust with scattering efficiencies similar to interstellar dust, then an analog to Hale-Bopp could
be 100 times brighter than an analog to the Earth. 

This work has been partly supported by NASA under grant NAG5-7924.

\newpage
\begin{figure}
\epsscale{1}
\plotone{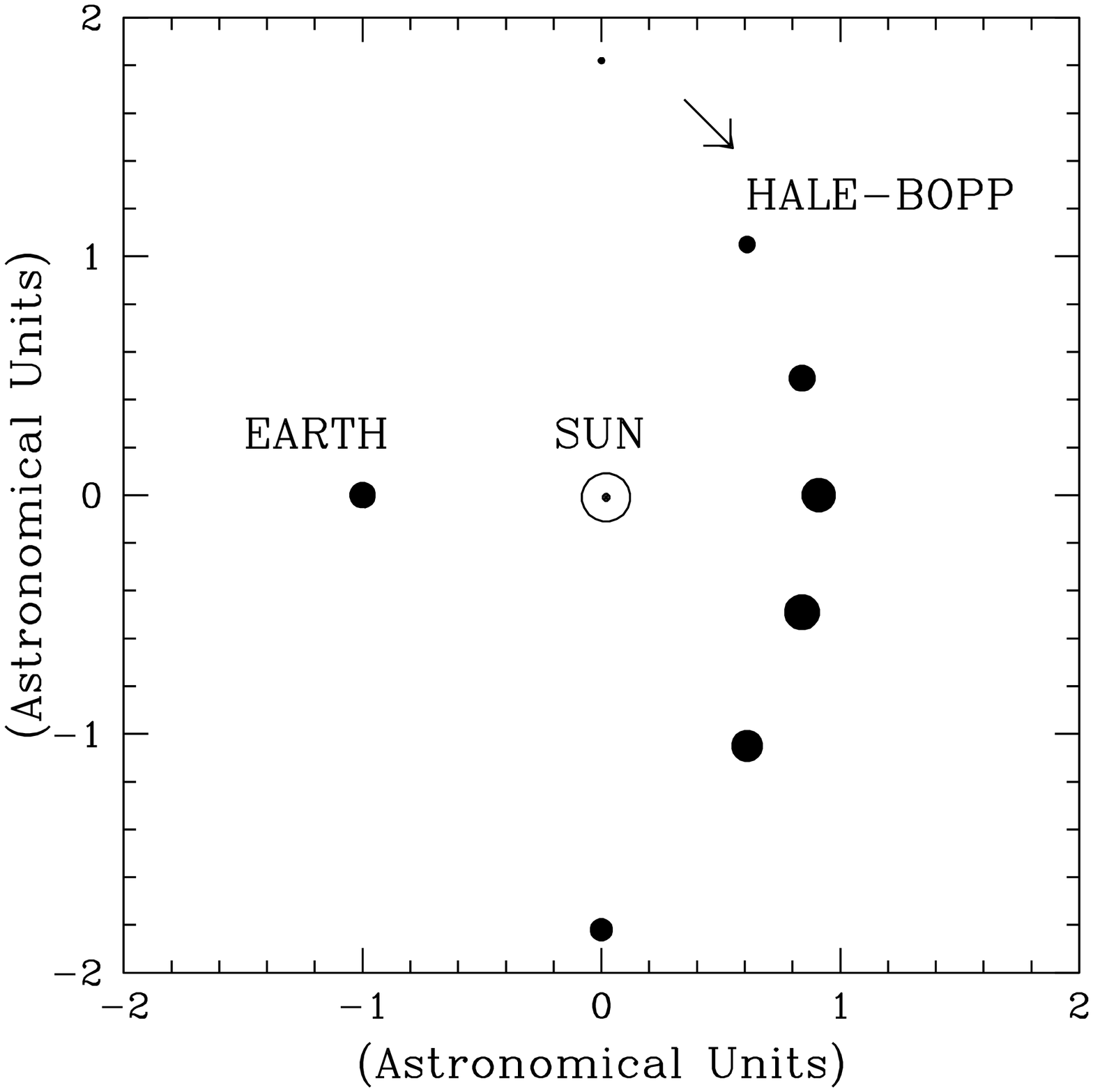}
\caption{The area of each solid point scales as the optical brightness of the image of our estimate for Hale-Bopp and the Earth.  We assume both the comet's and the planet's orbits are viewed face-on, and thus we only witness half of the illuminated hemisphere of the planet.  The comet enters near the top of the diagram and exits near the bottom.
}
\end{figure}
\begin{figure}
\epsscale{1}
\plotone{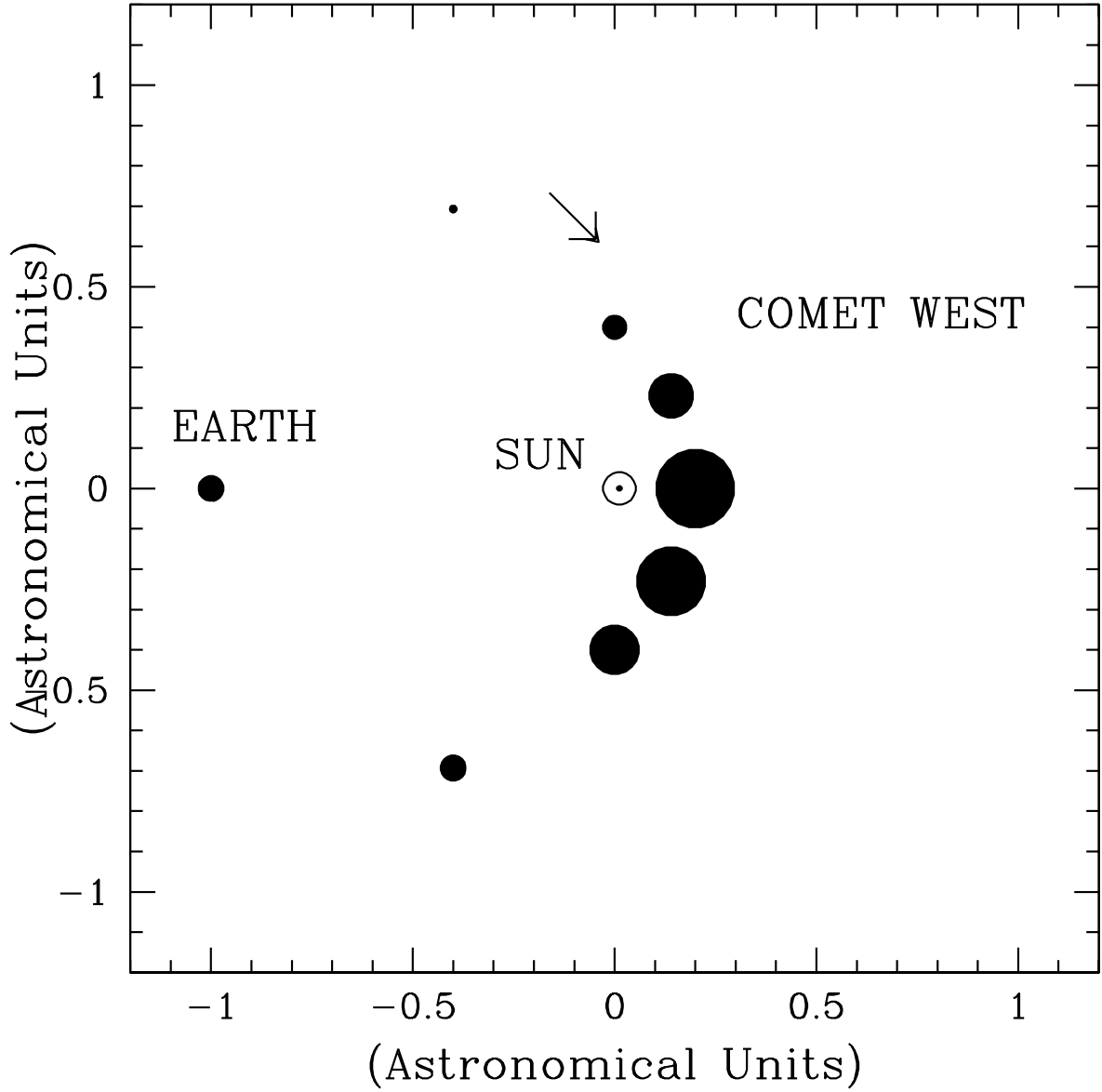}
\caption{The same as Figure 1 except that we show the results for an analog of Comet West.}
\end{figure}

\begin{figure}
\epsscale{1}
\plotone{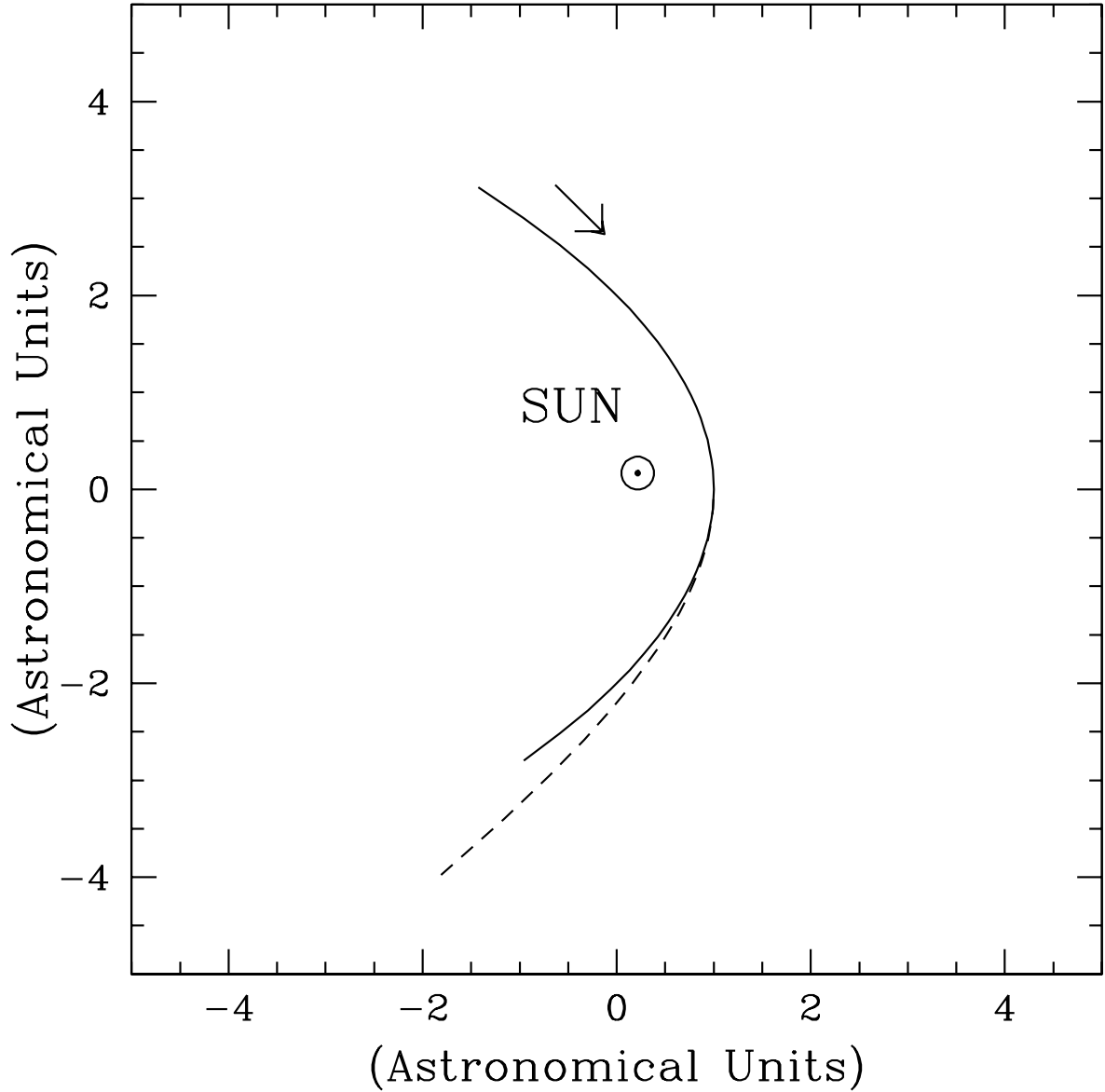}
\caption{The solid line shows the inner portion of an orbit for a comet with eccentricity of 0.9999 and $R_{p}$ = 1 AU while the dotted line shows
 the orbit of a dust grain with ${\beta}$ = 0.2 that is ejected from the comet at perihelion.    In the portions of the orbit where the comet is active, the distance
between the dust grain and the comet is significantly smaller than the comet's
distance from the Sun.}
\end{figure}
\begin{figure}
\epsscale{1}
\plotone{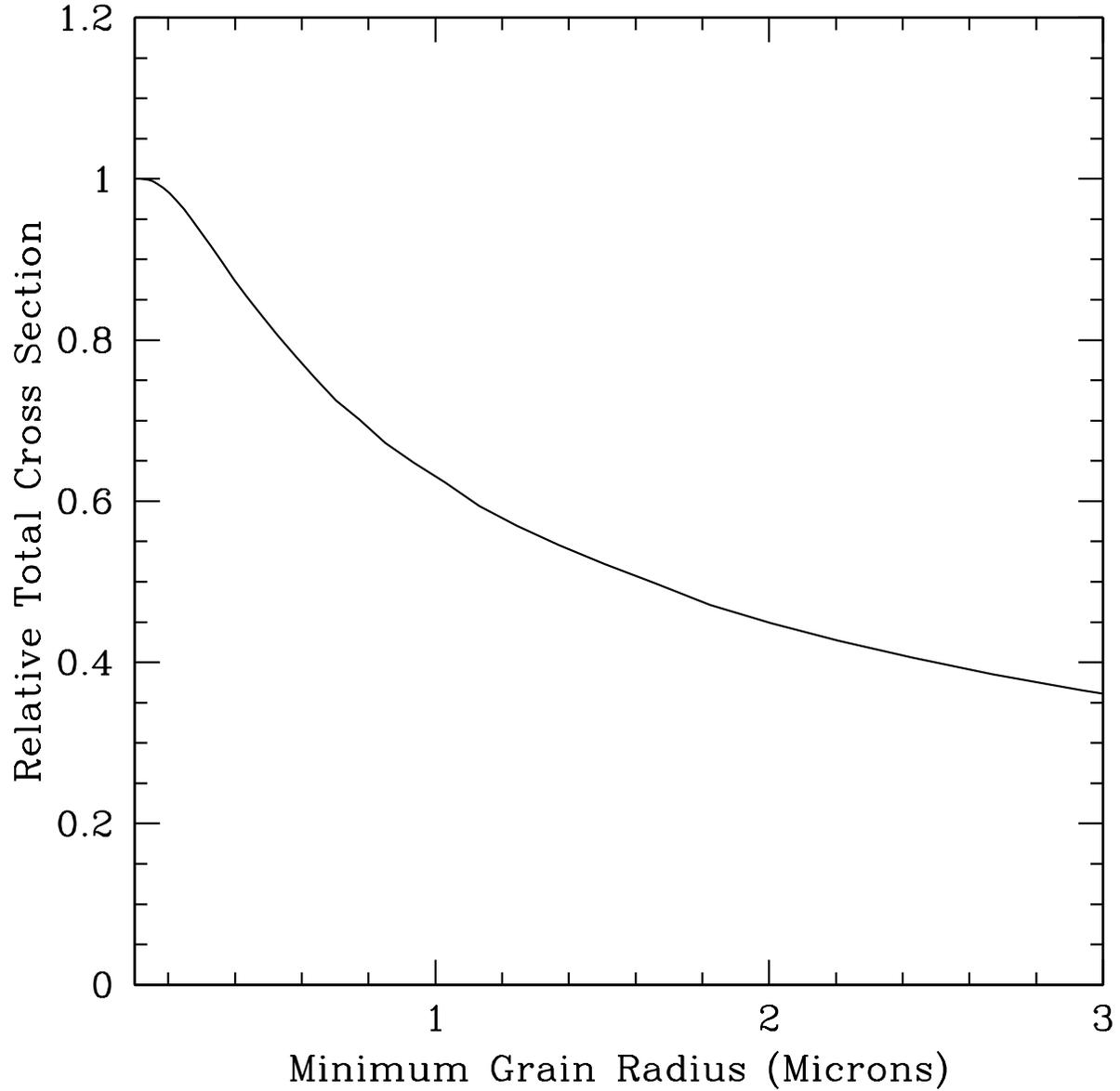}
\caption{Using equations (17) and (18), we show  ${\sigma}_{tot}(r_{min})/{\sigma}_{tot}(0.1\,{\mu}m)$ vs. $r_{min}$ to show how the integrated cross section diminishes as the minimum grain size is increased.}
\end{figure}\end{document}